\documentstyle[12pt]{article}

\newcommand{\EQ}{\begin{equation}}
\newcommand{\EN}{\end{equation}}
\newcommand{\bea}{\begin{eqnarray}}
\newcommand{\ena}{\end{eqnarray}}
\newcommand{\vs}[1]{\vspace{#1 mm}}
\newcommand{\hs}[1]{\hspace{#1 mm}}
\def\bbox{{\,\lower0.9pt\vbox{\hrule \hbox{\vrule height 0.2 cm
\hskip 0.2 cm \vrule height 0.2 cm}\hrule}\,}}

\newcommand{\sumi}{\sum_{i=1}^{N_0}}
\newcommand{\sumij}{\sum_{i,j=1 \atop i<j}^{N_0}}
\newcommand{\suma}{\sum_{a=1}^{N-1}}
\newcommand{\waa}[1]{\sum_{#1>0}}
\newcommand{\wab}[2]{\sum_{#1,#2>0}}

\renewcommand{\a}{\alpha}
\renewcommand{\b}{\beta}
\newcommand{\bp}{\beta'}
\renewcommand{\c}{\gamma}
\renewcommand{\d}{\delta}

\newcommand{\la}{\lambda}
\newcommand{\La}{\Lambda}

\newcommand{\pa}{\partial}

\newcommand{\dsl}{\pa \kern-0.5em /}
\newcommand{\nn}{\nonumber\\}

\newcommand{\dza}[1]{\frac{dz^a_{#1}}{2\pi i}}
\newcommand{\dzza}[1]{\frac{dz^a_{#1}}{2\pi i z^a_{#1}}}

\newcommand{\pr}[2]{\prod_{#1=1}^{#2}}
\newcommand{\prij}[1]{\prod_{i,j=1\atop i<j}^{#1}}
\newcommand{\factor}{\frac{2\pi}{L}}
\newcommand{\shalf}{\frac{1}{2}}

\newcommand{\hi}{\langle}
\newcommand{\mi}{\rangle}

\begin{document}
\topmargin 0pt
\oddsidemargin 5mm

\newcommand{\NP}[1]{Nucl.\ Phys.\ {\bf #1}}
\newcommand{\AP}[1]{Ann.\ Phys.\ {\bf #1}}
\newcommand{\PL}[1]{Phys.\ Lett.\ {\bf #1}}
\newcommand{\NC}[1]{Nuovo Cimento {\bf #1}}
\newcommand{\CMP}[1]{Comm.\ Math.\ Phys.\ {\bf #1}}
\newcommand{\PR}[1]{Phys.\ Rev.\ {\bf #1}}
\newcommand{\PRC}[1]{Phys.\ Rep.\ {\bf #1}}
\newcommand{\PRL}[1]{Phys.\ Rev.\ Lett.\ {\bf #1}}
\newcommand{\PTP}[1]{Prog.\ Theor.\ Phys.\ {\bf #1}}
\newcommand{\PTPS}[1]{Prog.\ Theor.\ Phys.\ Suppl.\ {\bf #1}}
\newcommand{\MPL}[1]{Mod.\ Phys.\ Lett.\ {\bf #1}}
\newcommand{\IJMP}[1]{Int.\ Jour.\ Mod.\ Phys.\ {\bf #1}}
\newcommand{\JP}[1]{Jour.\ Phys.\ {\bf #1}}
\newcommand{\JMP}[1]{Jour.\ Math.\ Phys.\ {\bf #1}}

\begin{titlepage}
\setcounter{page}{0}
\begin{flushright}
OU-HET 239 \\
hep-th/9603070
\end{flushright}

\vspace{1cm}
\begin{center}
{\Large 
EXACT SOLUTIONS OF GENERALIZED CALOGERO-SUTHERLAND MODELS \\
--- $BC_N$ and $C_N$ cases ---
}
\vspace{1.5cm}

{\large 
Masamichi Kojima\footnote{e-mail address:
 koji2@phys.wani.osaka-u.ac.jp}
and Nobuyoshi Ohta\footnote{e-mail address:
 ohta@phys.wani.osaka-u.ac.jp}}

\vspace{1cm}
{\em Department of Physics, Osaka University,\\
Toyonaka, Osaka 560, Japan}

\end{center}
\vspace{15mm}
\centerline{{\bf{Abstract}}}
\vspace{.5cm}

Using a collective field method, we obtain explicit solutions
of the generalized Calogero-Sutherland models that are characterized by
the roots of the classical groups $B_N$ and $C_N$. Starting from the
explicit wave functions for $A_{N-1}$ type expressed in terms of the singular
vectors of the $W_N$ algebra, we give a systematic method to construct
wave functions and derive energy eigenvalues for other types of theories.

\end{titlepage}

\newpage
\renewcommand{\thefootnote}{\arabic{footnote}}
\setcounter{footnote}{0}

The Calogero-Sutherland (CS) models~\cite{CS} describe one-dimensional
quantum systems of $N_0$ particles on a circle interacting
with each other by an inverse square potential. The complete excitation
spectrum and wave functions are exactly calculable in these models.
They play significant roles in various subjects such as fractional
statistics~\cite{HAL,HA}, quantum Hall effect~\cite{K} and $W_\infty$
algebra~\cite{HW}.

Although it had been difficult to solve this eigenvalue problem directly,
Stanley and Macdonald~\cite{SM} found that the solutions are expressed by
Jack symmetric polynomials and studied their properties. However,
they did not show how to construct Jack polynomials, which
are necessary to calculate correlation functions explicitly. Thus it
is an important problem to find a systematic method to construct
Jack polynomials.

Recently this problem has been solved by the use of collective field
method~\cite{COL} and conformal field theory technique by Awata
et al.~\cite{AMOS}.
They have shown that the Hamiltonian can be expressed in terms of
Virasoro and $W_N$ generators of positive modes and hence
the Jack symmetric polynomials can be represented as $W_N$ singular vectors,
whose explicit forms are given by integral representations using free bosons.

Among many variants of the CS models~\cite{HAL}, a class of models have
been known to be exactly solvable and show interesting behaviors similar to
the original ones. They are the Lie-algebraic generalization of the above
models~\cite{OP}. In particular the so-called CS model of $BC_N$-type
(hereafter referred to as $BC_N$-CS model) is the most general
one with $N_0$ interacting particles. This model is known to be
relevant to one-dimensional physics with boundaries. The energy eigenvalues
for these models have been obtained for both ground and excited
states~\cite{Y,BPS}, but the wave functions have been known only for
the ground states~\cite{OP,BPS}. The purpose of this paper is to give
a systematic method to construct the wave functions for excited states
explicitly and also give elementary derivation of the energy eigenvalues
by using the collective field method.

The Hamiltonian $H_{CS}$ for ordinary CS models is given by
\EQ
H_{CS}=-\sumi\frac{1}{2} \frac{\pa^2}{\pa q_i^2}
+\left(\frac{\pi}{L}\right)^2 \sumij
\frac{\b(\b-1)}{\sin^2\frac{\pi}{L}(q_i-q_j)},
\label{cs}
\EN
where $L$ and $\b$ are the circumference of the circle and a coupling
constant, respectively. The ground state $\Psi_0$ and the energy
eigenvalue $E_0$ of $H_{CS}$ is given by~\cite{CS,OP}
\bea
\Psi_0&\equiv&\Delta_{CS}^{\b} = \left(\frac{L}{\pi}\prod_{i,j=1 \atop i<j}
 ^{N_0}\sin\frac{\pi}{L}(q_i-q_j)\right)^{\b}, \nn
E_0&=&\frac{\b^2}{6}\left(\frac{\pi}{L}\right)^2(N_0^3-N_0).
\label{gene}
\ena
Note that the ground state exhibits fractional statistics for rational $\b$.
The excited states of $H_{CS}$ take the form
$\Psi=\Delta_{CS}^{\b} J_\la (q;\b)$. The functions $J_\la(q;\b)$ are 
known as the Jack polynomials characterized by an index $\la$ of the Young
diagram and are symmetric in the coordinates $q_i$ so that the statistics
of the system is determined by the wave function of the ground state.

Looking at the structure of the above Hamiltonian~(\ref{cs}), one
immediately recognizes that there is a close relation of this model to
the root system
of the classical group $A_{N-1}$. It is then natural to consider
Lie-algebraic generalization of this model. Indeed, it has been
known for some time~\cite{OP} that the models described by the
following Hamiltonian are exactly solvable:
\EQ
H_{GCS}=-\sumi\frac{1}{2}\frac{\pa^2}{\pa q_i^2}
 +\shalf\left(\frac{\pi}{L}\right)^2 \sum_{\vec{\a}\in R_{+}}
\frac{\mu_\a(\mu_\a+2\mu_{2\a}-1)|\vec{\a}|^2}
{\sin^2\frac{\pi}{L}(\vec{\a}\cdot\vec{q})},
\label{gcs}
\EN
where $R_+$ stands for positive roots of the classical group
under consideration and the coupling constants $\mu_\a$ are equal
for the roots of the same length. The most general model then
is the one with all the roots in $B_N$ and $C_N$ algebras.
This is the $BC_N$-CS model we are going to discuss.

To render our subsequent calculations simple, we introduce the following
variables:
\EQ
x_j\equiv\exp\left(\frac{2\pi i q_j}{L}\right) ; \qquad
D_i\equiv x_i\frac{\pa}{\pa x_i}.
\label{var}
\EN
Using these variables, the Hamiltonian (\ref{gcs}) is cast into
\bea
H_{GCS}&=& \shalf \left(\frac{2\pi}{L}\right)^2 \left[
 \sumi D_i^2-2\b(\b-1)\sumij  \left( \frac{x_i x_j}{(x_i-x_j)^2}
 +\frac{x_i x_j^{-1}}{(x_i-x_j^{-1})^2} \right) \right. \nn
& & \left. -\sumi  \left( \c(\c+2\d-1) \frac{x_i}{(x_i-1)^2}+4\d(\d-1)
 \frac{x_i^2}{(x_i^2-1)^2} \right) \right],
\label{hgcs}
\ena
where we have used $\b,\c,\d$ for coupling constants.
We note that putting $\c=0$ reduces the model to $C_N$-type,
$\d=0$ to $B_N$-type, and finally $\c=\d=0$ to $D_N$-type.
In this paper, we will be mainly concerned with solutions common to
all these root systems. The solutions for $B_N$ and $D_N$ contain
additional special ones corresponding to spinor representations,
which will be discussed in a separate paper.

In analogy to the solution~(\ref{gene}), the ground state wave
function and energy are given by~\cite{OP,BPS}
\bea
\Delta_{GCS} &=& \prod_{i=1}^{N_0} 
\left(\sin\frac{\pi}{L}q_i \right)^\c
\left(\sin\frac{2\pi}{L}q_i \right)^\d
\prod_{i,j=1 \atop i<j}^{N_0}
\left(\sin\frac{\pi}{L}(q_i-q_j)
\sin\frac{\pi}{L}(q_i+q_j) \right)^\b \nn
&=& \prod_{i=1}^{N_0} x_i^{-\b(N_0-1)-\c/2-\d}
 (x_i-1)^\c (x_i^2-1)^\d
 \prod_{i,j=1 \atop i<j}^{N_0} (x_i-x_j)^\b(x_ix_j-1)^\b \nn
E_0^{GCS} &=& \sumi \left[ \frac{\c}{2} + \d+\b(N_0-i)\right]^2.
\label{grou}
\ena

Now our task is to solve the eigenvalue problem
\EQ
H_{GCS} \Delta_{GCS} \Phi^{GCS} = E_{GCS} \Delta_{GCS} \Phi^{GCS}.
\EN
The effective Hamiltonian $H_{eff}$ acting on the function $\Phi^{GCS}(x)$
is derived by the transformation
\EQ
\Delta_{GCS}^{-1}H_{GCS}\Delta_{GCS}=\shalf\left(\factor
 \right)^2 \left[ H_{eff} + E_0^{GCS} \right],
\label{tran}
\EN
and our problem reduces to
\EQ
H_{eff}\Phi^{GCS} = E_{eff} \Phi^{GCS} \;\; ; \quad
E_{GCS} = \shalf \left(\frac{2\pi}{L}\right)^2
 \left[ E_0^{GCS} + E_{eff} \right].
\label{eig}
\EN
In terms of the variables (\ref{var}), we find
\bea
H_{eff}&=& \sumi D^2_i + \b \sumij \left(\frac{x_i+x_j}{x_i-x_j}
 (D_i-D_j) + \frac{x_i+x_j^{-1}}{x_i-x_j^{-1}}(D_i+D_j) \right) \nn
&& +\sumi \left( \c \frac{x_i+1}{x_i-1}
 +2\d  \frac{x_i+x_i^{-1}}{x_i-x_i^{-1}}  \right)D_i .
\label{ham}
\ena
We are going to express this Hamiltonian by free bosons.
In particular, it will be related to the free boson representation
of the $W_N$ algebra corresponding to the $A_{N-1}$
group.\footnote{Here $N$ is an arbitrary integer $(\geq 2)$
independent of $N_0$.}
Before jumping into this, let us first review relevant results in the
free boson representation of this algebra~\cite{AMOS}.

Let $\vec{e}_i$ $(i=1,\cdots, N)$ be an orthonormal basis ($\vec{e}_i
\cdot \vec{e}_j = \d_{ij}$). We define the weights of the vector
representation $\vec{h}_i$, the simple roots $\vec{\a}^a$ ($a=1,\cdots,
N-1$) and the fundamental weights $\vec{\La}_a$ by
\bea
&& \vec{h}_i = \vec{e}_i - \frac{1}{N}\sum_{j=1}^N \vec{e}_j, \qquad
\vec{\a}^a = \vec{h}_a-\vec{h}_{a+1}, \qquad
\vec{\La}_a = \sum_{i=1}^a \vec{h}_i, \nn
&& \vec{\a}^a\cdot \vec{\a}^b \equiv A^{ab} = 2\d^{a,b} - \d^{a,b+1}
-\d^{a,b-1}, \qquad
\vec{\a}^a\cdot \vec{\La}_b \equiv A^a_b = \d^a_b.
\ena
We then introduce $N-1$ free bosons
\EQ
\vec{\phi}(z)=\sum_{a=1}^{N-1}\phi^a(z)\vec{\La}_a
 =\sum_{a=1}^{N-1}\phi_a(z)\vec{\a}^a .
\EN
They have the mode expansion
\EQ
\vec{\phi}(z) = \vec{q} + \vec{a}_0 \ln z - \sum_{n \neq 0} \frac{1}{n}
\vec{a}_n z^{-n},
\EN
with the commutation relations
\EQ
[a_n^a,a_m^b]=A^{ab} n\d_{n+m,0}, \qquad
[a_0^a,q^b]= A^{ab},
\EN
The boson Fock space is generated by the oscillators of
negative modes on the state
\EQ
|\vec{\la} \mi = e^{\vec{\la}\cdot\vec{q}} |\vec{0}\mi; \qquad
\vec{a}_n |\vec{0}\mi = 0 \;\; (n\geq 0).
\EN
$\hi \vec{\la} |$ is similarly defined with the inner product
$\hi\vec{\la} |\vec{\la}' \mi=\d_{\vec{\la},\vec{\la}'}$.

The spin 2 and 3 generators of the $W_N$ algebra are given
by~\cite{FL}
\bea
T(z) &\equiv& \sum_n L_n z^{-n-2} \nn
&=& \shalf (\pa\vec{\phi}(z))^2
 + \a_0 \vec{\rho}\cdot\pa^2 \vec{\phi}, \nn
W(z) &\equiv& \sum_n W_n z^{-n-3} \nn
&=& \sum_{a=1}^{N-1} (\pa\phi_a(z))^2 \left( \pa\phi_{a+1}(z)
 - \pa\phi_{\a-1}(z) \right) \nn
&& + \a_0 \sum_{a,b=1}^{N-1} (1-a)A^{ab} \pa\phi_a(z)\pa^2\phi_b(z)
 + \a_0^2 \sum_{a=1}^{N-1} (1-a)\pa^3\phi_a(z),
\ena
where
\bea
\a_0 &=& \sqrt{\b}-\frac{1}{\sqrt{\b}}, \nn
\vec{\rho} &=& \sum_{a=1}^{N-1} \vec{\La}_a ; \;\;
(\vec{\rho})^2=\frac{1}{12}N(N^2-1).
\ena
The highest weight states of the $W_N$ algebra are created from
the vacuum by the vertex operator as
$|\vec{\la}\mi =:e^{\vec{\la}\cdot\vec{\phi}(0)}:|\vec{0}\mi$,
whose conformal weight $h(\vec{\la})$ and $W_0$-eigenvalue
$w(\vec{\la})$ are
\bea
h(\vec{\la}) &=& \shalf\left[ (\vec{\la}-\a_0 \vec{\rho})^2
 -\a_0^2(\vec{\rho})^2 \right], \nn
w(\vec{\la}) &=& \sum_{a=1}^{N-1} \left[
 \la_a^2(\la_{a+1} -\la_{a-1}) + \a_0 \left( 2(a-1)\la_a
 + (1-2a)\la_{a+1}\right) \la_a \right. \nn
&& \left. \hs{10} + 2\a_0^2(1-a)\la_a \right].
\label{wei1}
\ena
If we define $\vec{\la}^\pm_{\vec{r},\vec{s}}$ by
\bea
\vec{\la}^+_{\vec{r},\vec{s}} &=& \sum_{a=1}^{N-1}\left[
 (1+r^a-r^{a-1})\sqrt{\b}-(1+s^a)/\sqrt{\b}\right] \vec{\La}_a, \nn
\vec{\la}^-_{\vec{r},\vec{s}} &=& \sum_{a=1}^{N-1}\left[
 (1+r^a)\sqrt{\b}-(1+s^a-s^{a-1})/\sqrt{\b}\right] \vec{\La}_a,
\ena
singular vectors at level $\sum_{a=1}^{N-1}r^a s^a$ with the highest
weight $|\vec{\la}^\pm_{\vec{r},\vec{s}}\mi$ are given as
\bea
|\chi_{\vec{r},\vec{s}}^{+}\mi
&=&\oint\pr{a}{N-1}\pr{j}{r^a}
 \dza{j} :e^{\sqrt{\b}\phi^a(z_j^a)}:
 |\vec{\la}_{\vec{r},\vec{s}}^{+}-\sqrt{\b}\sum_{a=1}^{N-1}
 r^a\vec{\a}^a\mi\nn
&=&\oint\pr{a}{N-1}\pr{j}{r^a}\dzza{j}
 \pr{a}{N-1}\prij{r^a} \left( z_i^a-z_j^a \right)^{2\b}
 \pr{a}{N-1}\pr{i}{r^a}\pr{j}{r^{a+1}}
 \left( z_i^a-z_j^{a+1} \right)^{-\b} \nn
&&\times\pr{a}{N-1}\pr{i}{r^a} \left( z_i^a \right)^{
 \left( 1-r^a+r^{a+1} \right)\b -s^a}
 \pr{a}{N-1}\pr{i}{r^a} e^{\sqrt{\b}\phi_{-}^a (z_i^a)}
 |\vec{\la}_{\vec{r},\vec{s}}^{+}\mi,
\label{sing}
\ena
where $\phi_-^a$ is defined by
$\phi_-^a (z) = \sum_{n > 0} \frac{1}{n} a_{-n}^a z^n$.
One can define similar singular vectors $|\chi_{\vec{r},\vec{s}}^{-}\mi$
at the same level using $\vec{\la}^-_{\vec{r},\vec{s}}$.
These singular vectors are annihilated by Virasoro $L_n$ and $W_n$
generators of positive modes and correspond to the following Young diagrams
parameterized by the numbers of boxes in each
row, $\la=(\la_1,\cdots,\la_N)$, $\la_1\geq\cdots\geq\la_N\geq 0$:

\vs{2}
\noindent
\makebox[  4cm]{ }
\makebox[  2cm]{$s^1$}\hskip-.4pt
\makebox[1.7cm]{$s^2$}
\makebox[1.4cm]{ }
\makebox[1.4cm]{$s^{N-2}$}\hskip-.35pt
\makebox[1.3cm]{$s^{N-1}$}
\hfill\break
\makebox[  4cm][r]{$\hfill\la=$}
\framebox[  2cm][l]{\rule[  -1cm]{0cm}{  2cm}$r^1$}\hskip-.4pt
\framebox[1.7cm][l]{\rule[-0.7cm]{0cm}{1.7cm}$r^2$}
\makebox[1.4cm]   {\raisebox{.25cm}{$\cdots\cdots$}}
\framebox[1.4cm][l]{\rule[-0.4cm]{0cm}{1.4cm}\raisebox{.25cm}{$r^{N-2}$}
     }\hskip-.4pt
\framebox[1.3cm][l]{\rule[-0.2cm]{0cm}{1.2cm}\raisebox{.25cm}{$r^{N-1}$}}
\makebox[.5cm][r]{.}
\vs{3}

\noindent
We can read off the relation between $\la$ and $\vec{r},\vec{s}$ from
this diagram. We also note that
\EQ
h(\vec{\la}_{\vec{r},\vec{s}}^{+}-\sqrt{\b}\sum_{a=1}^{N-1} r^a\vec{\a}^a)
= h(\vec{\la}_{\vec{r},\vec{s}}^{+}) + \sum_{a=1}^{N-1} r^a s^a.
\label{wei2}
\EN

Let us give a simple example at level 3 for $\vec{r}=(2,1,0,\cdots,0);\;\;
\vec{s}=(1,1,0.\cdots,0)$ which corresponds to the Young diagram
\raisebox{-.61ex}{\shortstack{\framebox[3mm]{} \\[-1.1mm] \framebox[3mm]{}}}
\hs{-1.5}\raisebox{.7ex}{\framebox[3mm][l]{}} \hs{-2} .
From (\ref{sing}), we find the singular vector is given by
\EQ
|\chi\mi \propto \left[ a_{-3}^1+\frac{\b-1}{\sqrt{\b}}a_{-1}^1 a_{-2}^1
 -(a_{-1}^1)^3 -\frac{\b+2}{2\b} \left((a_{-1}^1)^2
 -\sqrt{\b} a_{-2}^1 \right) a_{-1}^2 \right] |\la\mi.
\EN

Coming back to our problem, since our system~(\ref{ham}) has the
reflection invariance under $x_i \to x_i^{-1}$ for each $i$ in addition
to the permutation symmetry $x_i \leftrightarrow x_j$,
we expect that the solutions are invariant under these transformations.
It is then natural to define the symmetric power sums\footnote{
In the $A_{N-1}$ case, we define $p_n=\sumi x_i^n$. Eq.~(\ref{power})
is the generalization necessary in our system where we do not have
translational invariance.}$^,$\footnote{It is known in mathematical
literature that the representation ring for $BC_N$ and $C_N$ systems
is isomorphic to the ring generated by these symmetric power sums.
For the special cases of spinor representations in $B_N$ and $D_N$,
we need additional functions $\prod_{i=1}^{N_0}( \sqrt{x_i}
+ 1/\sqrt{x_i} )$ for $B_N$ and $\prod_{i=1}^{N_0}(\sqrt{x_i}
- 1/\sqrt{x_i} )$ as well for $D_N$. These special cases will be
discussed in a separate paper.}
\EQ
p_n=\sumi (x_i^n+x_i^{-n}),
\label{power}
\EN
in terms of which we can express our effective Hamiltonian.
We then consider the map
\bea
| f\mi\mapsto f(x)&\equiv&\hi\vec{\la}| C_{\bp}| f\mi \nn
C_{\bp} &\equiv& \exp \left(\bp\waa{n}\frac{1}{n}a_{n,1}p_n\right),
\label{map}
\ena
where $\bp$ is a parameter to be determined shortly. This gives the
following correspondence between the oscillators and the power sums:
\EQ
\bp p_n \leftrightarrow a_{-n}^1 ;\;\;
\frac{n}{\b'}\frac{\pa}{\pa p_n} \leftrightarrow a_{n,1} .
\label{rule}
\EN
Note that $a_{-n}^a \; (a>1,n>0)$ vanishes under this map.

With the help of the map (\ref{map}), the Hamiltonian (\ref{ham})
is transformed into\footnote{It appears that this Hamiltonian (\ref{ham1})
is not Hermitian. This is simply because we have transformed our
Hamiltonian by the ground state (see eq.~(\ref{tran})).}
\bea
\hat{H}_{eff} &=& \wab{n}{m}\left(
 \bp a_{-n-m}^1 a_{n,1} a_{m,1}+\frac{\b}{\bp}a_{-n}^1 a_{-m}^1 a_{n+m,1}
      \right)\nn
&&+\waa{n}\left[ n(1-\b)+\left\{ \b(2N_0-1)+\c+2 \d \right\}
  \right]a_{-n}^1 a_{n,1} \nn
&&-2 \bp\wab{n}{m}a_{-m}^1 a_{n,1}a_{n+m,1}
-2\wab{n}{m}\left\{
  (\b-2\d)a_{-m}^1 a_{2n+m,1}-\c a_{-m}^1 a_{n+m,1} \right\}\nn
&&-2\bp N_0 \waa{n}\left\{
  (\b-2\d)a_{2n,1}-\c a_{n,1}+\bp (a_{n,1})^2  \right\}.
\label{ham1}
\ena
Here and in what follows, carets on the Hamiltonian and states mean
that they are expressed in terms of oscillators.
After straightforward calculation, one finds that this Hamiltonian
can, for the choice
\EQ
\bp=\sqrt{\b},
\EN
be finally rewritten as
\bea
\hat{H}_{eff} &=& \hat{H}' + \sum_{n>0} \hat{H}_{n}
 + \sum_{a>1}\waa{n} a_{-n}^a (\cdots) \nn
&& + \sqrt{\b}\waa{n}\left(\frac{2}{N_0}a_{-n}^1 L_n-2a_{n,1}L_n\right)
+2\waa{n}\left\{\c L_n+\left(2 \d-\b\right)L_{2n}\right\},
\label{res}
\ena
where
\bea
\hat{H}' &=& \waa{n}\vec{a}_{-n}\cdot\vec{a}_n \left( 2N_0 \b-1+\c+2\d
 - 2\sqrt{\b}a_{0,1} \right) +\sqrt{\b}\left(W_0-W_{0,zero}\right), \nn
\hat{H}_{n} &=& 2\c \suma \sum_{m=1}^{n-1}a_{n-m,a}
 \left( a_{m,a+1}-a_{m,a} \right)+\frac{2\c}{\sqrt{\b}} a_{n,1}
 \left\{ (n+1)(\b-1)+ N_0 \b-\sqrt{\b}a_0^1 \right\} \nn
&&+\frac{2\c}{\sqrt{\b}}\sum_{a=2}^{N-1}a_{n,a}
 \left\{ (n+1)(\b-1)-\sqrt{\b}a_0^a \right\} \nn
&& + 2\sqrt{\b}\suma \sum_{m=1}^{n-1}
 a_{n,1}( a_{n-m,a}a_{m,a}-a_{n-m,a+1}a_{m,a} ) \nn
&& + 2(\b-2\d)\suma \sum_{m=1}^{2n-1}a_{2n-m,a}(a_{m,a}-a_{m,a+1}) \nn
&& - 2 (a_{n,1})^2 \left\{ (n+1)(\b-1)+N_0 \b -\sqrt{\b}a_0^1 \right\} \nn
&&-2\sum_{a=2}^{N-1} a_{n,1}a_{n,a}
 \left\{ (n+1)(\b-1)-\sqrt{\b}a_0^a \right\} \nn
&&-\frac{2}{\sqrt{\b}} \left[ (\b-2\d)a_{2n,1} \left\{
 (2n+1)(\b-1)+N_0 \b-\sqrt{\b}a_0^1 \right\} +\b a_{2n,1} \right]\nn
&&-\frac{2}{\sqrt{\b}}(\b-2\d)\sum_{a=2}^{N-1}a_{2n,a}
 \left\{ (2n+1)(\b-1)-\sqrt{\b}a_0^a \right\}.
\label{res1}
\ena
Here $W_{0,zero}$ in $\hat{H}'$ is the zero mode part of $W_0$.
The third term involving $a_{-n}^a \; (a>1,n>0)$ in (\ref{res})
vanishes after multiplying by $\hi \vec{\la}|C_{\bp}$ and will be
disregarded in the following. Note that $\hat{H}'$ is the sum
of number operators and $W_N$ zero mode and also that
$\hat{H}_{n}$ consist of annihilation operators only.

To construct our eigenstates of the Hamiltonian $\hat{H}_{eff}$, consider
singular vectors at the level $\sum_{a=1}^{N-1}r^a s^a$. Since these are
annihilated by Virasoro generators $L_n$ of positive modes,
only the first two terms in (\ref{res}) are relevant to our problem.
These are already eigenstates of $\hat{H}'$ with the eigenvalue
\bea
E_{\la}&=& \left[ h\left(\vec{\la}^+_{\vec{r},\vec{s}}
 - \sqrt{\b}\sum_{a=1}^{N-1}r^a\vec{\a}^a\right)
 - h\left(\vec{\la}^+_{\vec{r},\vec{s}} \right)\right]
 \left[ 2N_0 \b-1+\c+2\d - 2 \left( \b r_1-s_1+\sqrt{\b}\a_0\rho_1\right)
 \right] \nn
&& + \sqrt{\b} \left[ w\left(\vec{\la}^+_{\vec{r},\vec{s}}
 - \sqrt{\b}\sum_{a=1}^{N-1}r^a\vec{\a}^a\right) -
w\left(\vec{\la}^+_{\vec{r},\vec{s}} \right) \right] \nn
&=& \sum_{a=1}^{N-1} r^a s^a s^a+2\sum_{a,b=1\atop a>b}^{N-1}r^a s^a s^b
 +\sum_{a=1}^{N-1} r^a s^a (2N_0 \b-\b+\c+2\d-\b r^a), \nn
&=& \sumi \left[\la_i^2+2  \left\{ \b(N_0-i)+\frac{\c}{2}+\d
 \right\}\la_i \right].
\label{exen}
\ena
Here use has been made of eqs.~(\ref{wei1}) and (\ref{wei2}) in deriving
the second equality, and of the relation between $\la$ and $\vec{r},\vec{s}$
obtained from the Young diagram in getting the third equality.

It is clear that applying $\hat{H}_{n}$ on the singular vectors produces
only states at the lower levels, and that the excitation energy is
given by the eigenvalue given in (\ref{exen}); $E_{eff}=E_\la$.
Thus the eigenstates of our system can be written as
\EQ
\hat{\Phi}_{\la}^{GCS}=\hat{J}_{\la}+\sum_{\mu<\la}C_{\mu}\hat{J}_{\mu},
\label{wvf}
\EN
where $\hat{J}_\la$ is the oscillator representation of the Jack polynomials
for the $A_{N-1}$ case (or the $W_N$ singular vectors) with the coefficients
$C_\mu$ to be determined from the highest state $\hat{J}_\la$ by the
application of $\hat{H}_n$.

To be more explicit, by taking the inner product of eq.~(\ref{wvf}) with
$\hi J_\nu|\hat{H}_{eff}$, we have
\EQ
\hi J_{\nu}|\waa{n}\hat{H}_{n}|J_{\la}\mi
+\sum_{\mu<\la}C_{\mu}\hi J_{\nu}|\waa{n}\hat{H}_{n}|J_{\mu}\mi
= C_{\nu} \left( E_{\la}-E_{\nu} \right), \quad
 \left( \mbox{}\nu<\la \right),
\label{mas}
\EN
which is the master equation to determine the coefficients
$C_\mu$ successively.\footnote{It is easy to check from (\ref{exen})
that the energy difference $E_\la-E_\nu>0$ for $\la>\nu$. Hence the
energy denominator never vanishes in our eq.~(\ref{mas}).}
The inner product is easily evaluated by using the oscillator
representation given in (\ref{res1}).
For example, choosing $\nu=\la-1$ in (\ref{mas}), the second coefficient
is found to be
\EQ
C_{\la-1}=\frac{\hi J_{\la-1}| \hat{H}_1 |J_{\la}\mi}{E_{\la}-E_{\la-1}},
\EN
where $\la-1$ stands for the Young diagram with single box removed from
$\la$. Next, setting $\nu=\la-2$, we get an equation for $C_{\la-2}$
involving only $C_{\la-1}$, which is already known. In this way, all
the coefficients can be obtained from (\ref{mas}) successively.
 
The actual eigenstates in terms of the symmetric power sums
(\ref{power}) can be read off from the explicit expression in
terms of the boson oscillators by the rule~(\ref{rule}).
The total energy is obtained from (\ref{grou}) and (\ref{exen}) as
\EQ
E_0^{GCS}+E_{eff}=\sumi \left[\la_i+\b(N_0-i)+\frac{\c}{2}+\d\right]^2 ,
\EN
in agreement with the known results~\cite{BPS}.

This completes our procedure to determine the whole eigenfunctions
and our elementary derivation of the excitation energy.

As a simple example of the application of our method, we present
the results of the eigenstates. For simplicity, we list only those
for the $D_N$ case ($\c=\d=0$) up to level 3:

{\bf level 1}
\EQ
p_1=J_{\vec{r}=(1,0,\cdots)\atop \vec{s}=(1,0,\cdots)} \;\; ; \;\;
\framebox[3mm]{} \;\; ; \;\; 1+2\b(N_0-1).
\EN

{\bf level 2}
\bea
&& p_2-p_1^2-4N_0 (\b-1)
 =J_{\vec{r}=(2,0,\cdots)\atop \vec{s}=(1,0,\cdots)}-4N_0 (\b-1)
\;\; ; \;\;
\shortstack{\framebox[3mm]{} \\[-1.1mm] \framebox[3mm]{}}
\;\; ; \;\; 2(1-3\b+2N_0\b), \nn
&& p_2+\b p_1^2-8N_0 \b
 =J_{\vec{r}=(1,0,\cdots)\atop \vec{s}=(2,0,\cdots)}-8N_0 \b
\;\; ; \;\; 
\framebox[3mm]{}\framebox[3mm][l]{}
\;\; ; \;\; 4(1-\b+N_0 \b).
\ena

{\bf level 3}
\bea
&& p_3-\frac{3}{2}p_1 p_2+\shalf p_1^3
 +\frac{3(N_0 -1)(\b-1)}{2N_0 \b-5\b+1}p_1\nn
&& \mbox{ }=J_{\vec{r}=(3,0,\cdots)\atop \vec{s}=(1,0,\cdots)}
 +\frac{3(N_0 -1)(\b-1)}{2N_0 \b-5\b+1}
 J_{\vec{r}=(1,0,\cdots)\atop \vec{s}=(1,0,\cdots)} \;\; ; \;\;
\shortstack{\framebox[3mm]{} \\[-1.1mm] \framebox[3mm]{}\\[-1.1mm]
\framebox[3mm]{}}
\;\; ; \;\; 3(1-4\b+2N_0 \b), \nn
&& p_3+(\b-1)p_1 p_2-\b p_1^3
 +\frac{2N_0 \b^2-8N_0 \b+5\b-2}{2N_0 \b-3\b+2}p_1 \nn
&& \mbox{ }= J_{\vec{r}=(2,1,0,\cdots)\atop \vec{s}=(1,1,0,\cdots)}
 +\frac{2N_0 \b^2-8N_0 \b+5\b-2}{2N_0 \b-3\b+2}
 J_{\vec{r}=(1,0,\cdots)\atop \vec{s}=(1,0,\cdots)} \;\; ; \;\;
\raisebox{-.61ex}{\shortstack{\framebox[3mm]{} \\[-1.1mm] \framebox[3mm]{}}}
\raisebox{.7ex}{\framebox[3mm][l]{}}
\;\; ; \;\; 5-8\b+6N_0 \b, \\
&& p_3+\frac{3}{2}\b p_1 p_2+\frac{\b^2}{2}p_1^3
 -\frac{3\b(N_0 \b+1)}{N_0 \b-\b+2}p_1 \nn
&& \mbox{ }= J_{\vec{r}=(1,0,\cdots)\atop \vec{s}=(3,0,\cdots)}
 -\frac{3\b(N_0 \b+1)}{N_0 \b-\b+2}
J_{\vec{r}=(1,0,\cdots)\atop \vec{s}=(1,0,\cdots)}
\;\; ; \;\;
\framebox[3mm]{} \framebox[3mm]{} \framebox[3mm][l]{}
\;\; ; \;\; 3(3-2\b+2N_0 \b). \nonumber
\ena
Here the corresponding Young diagrams and excitation energies are
also exposed.
 
To summarize, we have given a systematic algorithm to
compute eigenfunctions for excited states for the
$BC_N$-CS models. Remarkably these can be easily
obtained from those for the $A_{N-1}$ case (but modified to
be reflection invariant), which are nothing but singular
vectors of the $W_N$ algebra. Our method uses simple
oscillator representation, which is easily accesible for
physicists.

\section*{Acknowledgments}

We would like to thank K. Higashijima and H. Suzuki for valuable discussions.
Thanks are also due to H. Awata, Y. Matsuo, S. Odake and T. Yamamoto
for useful comments. Critical comments by D. Serban on the spinor
representations for $B_N$ and $D_N$ cases are also appreciated.

\end{document}